\documentstyle[prl,aps,psfig]{revtex}

\newcommand{\bea}{\begin{eqnarray}}
\newcommand{\eea}{\end{eqnarray}}
\newcommand{\beq}{\begin{equation}}
\newcommand{\eeq}{\end{equation}}

\newcommand{\marke}[1]{ \protect\label{#1}
}
\def\gsim{\mathrel{\raise.3ex\hbox{$>$\kern- .75em \lower1ex\hbox{$\sim$}}}}

\newfont{\cms}{cmss8 scaled 1440}

\def\/{\over}
\begin{document}

\parindent=1 em
\frenchspacing

\begin{center} {\LARGE \bf Pair correlation function of
 an inhomogeneous interacting Bose-Einstein condensate}
\\[1cm]
{\bf 
Markus Holzmann\footnote{
e-mail: holzmann@lps.ens.fr; castin@physique.ens.fr}
and Yvan Castin$\,^*$
\\[0.3cm]
\normalsize \it Laboratoire Kastler-Brossel
\footnote{Laboratoire Kastler Brossel is a unit\'e de recherche de
l'Ecole Normale Sup\'erieure et de l'Universit\'e Pierre et Marie Curie,
associ\'ee au CNRS.}\\
\normalsize \it and\\
\normalsize \it
CNRS-Laboratoire de Physique Statistique\\
\normalsize \it de l'Ecole Normale Sup\'erieure;\\
\normalsize \it 24, rue Lhomond;\\
\normalsize \it F-75005 Paris; France
}
\date{29 Jul 1998}
\vspace{0.2cm}
\end{center}
\begin{abstract}
We calculate the pair correlation function of an interacting
Bose gas in a harmonic trap
directly via Path Integral Quantum Monte Carlo simulation
for various temperatures and compare the numerical result with
simple approximative treatments. Around the critical temperature
of Bose-Einstein condensation, a description based on the 
Hartree-Fock approximation is found to be accurate.
At low
temperatures the Hartree-Fock approach fails and
we use a local density approximation based on
the Bogoliubov description for a homogeneous gas.
This approximation
agrees with 
the simulation results at low temperatures, where
the contribution of the phonon-like modes
affects the long range behavior of the correlation function.
Further we discuss the relation
between the pair correlation and quantities measured in recent
experiments.

\end{abstract}

\vspace{0.3cm}
\section{Introduction}
One of the appealing features of the experimental achievement of 
Bose-Einstein condensation in dilute vapors
\cite{anderson95,davis95,bradley95},
is the demonstration
of first order coherence of matter waves
\cite{andrews97}.
The interference pattern of this experiment agrees with 
the theoretical calculation
\cite{roehrl97}, which reveals
that the underlying theoretical concept of off-diagonal long range order
due to a macroscopically 
occupied quantum state is justified \cite{yang62}.
Additional experiments have explored certain aspects of second and third
order coherence of a trapped Bose gas \cite{ketterle97,burt97,shlyap85}.
Here we study the density-density correlation function
which is related to second order coherence.
With the knowledge of this pair correlation function,
the total interaction
energy can be calculated. In \cite{ketterle97} the release energy
of the atoms was measured after switching off the magnetic trap.
In the Thomas Fermi regime at zero
temperature the initial kinetic energy can be neglected
and the release energy is dominated by the interaction energy. By 
comparison with the usual mean field interaction energy using a 
contact potential, it was concluded that the release energy is mainly
proportional to the pair correlation function at vanishing
relative distance.
Strictly speaking this statement cannot be correct as for
interactions with a repulsive hard core the pair correlation
function must vanish at zero distance. To give a precise meaning
to this statement one needs to access the whole correlation function.

In this paper we consider in detail
the spatial structure of the correlation function
of an interacting trapped Bose gas. The 
Fourier transform of this function is directly related to the
static structure factor
which can be probed by off-resonant light scattering.
The 
tendency of bosonic atoms to cluster together
causes atom-bunching for an ideal
gas above the condensation temperature, 
for the atoms separated by less than the thermal
de-Broglie wavelength \cite{vanhove54}. For the condensate atoms,
this bunching vanishes,
since they all occupy the same quantum state \cite{london42,lemmens}. 
However, for a gas with strong repulsive interatomic interaction,
it is impossible
to find two atoms at exactly the same place,
and hence the pair correlation
function must vanish at very short distances.
This mutual repulsion can significantly
reduce the amount of bosonic bunching at temperatures around the 
transition temperature \cite{martin}. At much lower temperature,
the presence
of the condensate  changes the excitation spectrum as compared to the
noninteracting case.
It is known that in a homogeneous
Bose gas the modes of the phonons give rise to a modification of the
long range behavior of the correlation function \cite{huang}.

Using path integral quantum Monte Carlo simulations all equilibrium 
properties of Bose gases can be directly computed without any essential 
approximation \cite{Ceperley}. It has been shown that this calculation
can be performed directly for the particle numbers and temperatures of
experimental interest \cite{Werner}.
Here, we use this approach to calculate
the
pair correlation function  
for various temperatures and compare our results with simple
approximate treatments.

Near the critical temperature 
our data are quantitatively well explained by an 
improved
semiclassical Hartree-Fock theory,
where the full short range behavior is taken into account.
At low temperature this single-particle approximation
fails since the low lying energy modes become important and they are
not correctly described by the Hartree-Fock treatment.
In the Bogoliubov approach these modes are phonon-like and
change the behavior of the
correlation function.
Adapting the homogeneous Bogoliubov solution
locally to the inhomogeneous
trap case we find an excellent agreement with the Monte Carlo
simulation results at low temperature.

\section{Hamiltonian of the problem}
The Hamiltonian of $N$ interacting particles in an isotropic harmonic
trap with frequency $\omega$ is given by
\beq
H=\sum_{i=1}^{N} \left[ \frac{p_i^2}{2m}
 + \frac{1}{2}m\omega^2 r_i^2\right] +
\frac{1}{2}\sum_{i\neq j} V(r_{ij}),
\marke{hamilton}
\eeq
where $V$ is the interatomic potential, which depends only on the
relative distance $r_{ij}=|\vec{r}_i-\vec{r}_j|$ between two particles.
This potential in the experiments with alkali
atoms has many bound states, so that the Bose-condensed gases
are metastable systems rather than systems at thermal equilibrium.
To circumvent this theoretical difficulty, we have to replace the true
interaction potential by a model potential with no bound
states. 

This model potential is chosen in a way that it has the
same low energy binary scattering properties as the true interaction
potential. In the considered experiments, the $s$-wave contribution 
strongly dominates in a partial wave expansion of
the binary scattering problem, so that it is sufficient that
the model potential have the same $s$-wave scattering length 
$a$ as the true potential.
For simplicity we take in the quantum Monte Carlo
calculations a pure hard-core potential with diameter
$a$. 
In the analytical approximations of this paper, we have
taken, as commonly done in the literature,
the pseudo-potential described in \cite{huang}, which
is a regularized form of the contact potential, $g
\delta(\vec{r}_1-\vec{r}_2) \frac{\partial}{\partial r_{12}}
(r_{12} \cdot )$, with a coupling constant
\beq
g={4\pi\hbar^2a\over m}.
\eeq

\section{Path Integral Quantum Monte Carlo Approach}
\subsection{Reminder of the Method}
The partition function $Z$ of the system with inverse temperature
$\beta=(k_B T)^{-1}$ is given as the trace over the (unnormalized)
density matrix $\varrho$:
\beq
\varrho(\beta)=e^{-\beta H}
\eeq
over all symmetrized states.
Both satisfy the usual
convolution equation which we can write in the position representation:
\bea
Z & = &\frac{1}{N!}\sum_P \int \, d^{3N}R \, \varrho(R,R^P,\beta)\\
& = & \frac{1}{N!}\sum_P \int \, d^{3N}R \, \int \, d^{3N}R_2 \, ...
\int  \, d^{3N}R_M \, \varrho(R,R_2,\tau) ... \varrho(R_M,R^P,\tau).
\marke{conv}
\eea
Here $\tau=\beta/M$, where $M$ is an arbitrary integer,
$R$ is the 3N-dimensional vector of the particle
coordinates $R=(\vec{r}_1,\vec{r}_2,...,\vec{r}_N)$,
$P$ is a permutation of the $N$ labels of the atoms
and $R^P$ denotes the vector with
permuted labels:
$R^P=(\vec{r}_{P(1)},\vec{r}_{P(2)},...,\vec{r}_{P(N)})$.
Since only density matrices at higher temperature ($\tau\ll \beta$)
are involved, high temperature approximations of the 
$N$-body density matrix can be used.

The simplest approximation is the primitive approximation
corresponding to 
$\exp[\tau(A+B)] \simeq \exp[\tau B/2] \exp[\tau A]
\exp[\tau B/2]$, which neglects
the commutator of the operators $A$ and $B$. It corresponds to a 
discrete approximation of the Feynman-Kac
path integral and gives the correct result in the limit
$M \rightarrow \infty$
\cite{Feynman,Ceperley}. This can be seen by using the 
Trotter formula for the exponentials of a sum of two noncommuting
operators
\beq
e^{\tau(A+B)}=\lim_{n\to \infty} \left( 
e^{\tau A/n} e^{\tau B/n} \right)^n.
\eeq
The discretisized path integral for the $N$-particle density
matrix at inverse temperature $\tau$ can therefore
be written 
in the primitive approximation  with symmetric splitting as
\beq
\varrho(R,R',\tau)\simeq
\prod_{k=1}^N \varrho_1(\vec{r}_k,\vec{r}_k\,',\tau)
\prod_{i < j} \exp \left[ -\frac{\tau}{2}\left( V(\vec{r}_{ij} )
+ V(\vec{r}_{ij}\,') \right) \right],
\marke{simple}
\eeq
where $\varrho_1(\vec{r}_k,\vec{r}_k\,',\tau)$ is the
density matrix of noninteracting particles in the harmonic
trap and  $\vec{r}_{ij}=\vec{r}_i-\vec{r}_j$,
$\vec{r}\,'_{ij}=\vec{r}\,'_i-\vec{r}\,'_j$.
However, this approximation 
leads to slow convergence since the potential energy in the
argument of the exponentials are not slowly varying compared
to the density matrix of one particle in the external
potential, $\varrho_1(\vec{r}_i,\vec{r}_i\,',\tau)$.
This has the consequence that eq.(\ref{simple}) is not a smooth
function in the region where two particles are in contact, as
it should. In order to  get such a smooth function we use the fact
that the potential energy part of eq.(\ref{simple})
can also be written as:
\beq
e^{- \tau\left( V(\vec{r}_{ij} ) +V(\vec{r}_{ij}\,')\right)/2
} \simeq
g_{2}(\vec{r}_{ij};\vec{r}_{ij}\,',\tau)
=\left\langle e^{-\int_0^{\tau} d \, t \, V(\vec{r}_{ij}(t))
} \right\rangle_{rw},
\eeq
where the brackets correspond an average over an arbitrary 
distribution of $\vec{r}_{ij}(t)$,
starting from $\vec{r}_{ij}$ and
ending at $\vec{r}_{ij}\,'$, which reproduces the correct 
high temperature limit of the primitive approximation.
It is convenient to take the random walk corresponding to the
kinetic energy as weight function so that $g_{2}$
is the solution of the binary scattering problem
in free space:
\beq
g_{2}(\vec{r}_{ij},\vec{r}\,'_{ij};\tau)=
\frac{\langle \vec{r}_{ij}| 
\exp[-\tau (p_{ij}^2/m
+ V(r_{ij}) )] | \vec{r}\,'_{ij} \rangle }{ \langle \vec{r}_{ij}|
 \exp[-\tau p_{ij}^2/m]
|  \vec{r}\,'_{ij} \rangle },
\eeq
where $\vec{p}_{ij}$ is the
operator of the relative momentum between particles $i$ and $j$.
This leads to the so called pair-product approximation
\cite{barker79,Ceperley}, where
the density matrix is approximated as
\beq
\varrho(R,R',\tau)\simeq
\prod_{n=1}^N \varrho_1(\vec{r}_n,\vec{r}_n\,',\tau)\prod_{i <
j}
g_{2}(\vec{r}_{ij};\vec{r}_{ij}\,',\tau).
\eeq
This approximation has the advantage to include exactly all
binary collisions of atoms
in free space, only three and more atoms in close proximity will
lead to an error; convergency with respect to $M \rightarrow \infty$ 
is reached much faster. 
In the simulation the two-particle correlation function $g_{2}$
is equal to one for non-interacting particles
and plays the role of
a binary correction term in presence of two-body interactions.

As in \cite{Werner} we take $N=10,000$ particles with a hard-core
radius of $a=0.0043 (\hbar/m\omega)^{1/2}$. The transition
temperature of the noninteracting Bose-gas is $k_B T_c^0 =
20.26\, \hbar \omega$ or $\beta_c^0 \simeq 0.05 (\hbar \omega)^{-1}$ and
a value of $\tau=0.01 (\hbar \omega)^{-1}$ was found sufficient.
In the low
temperature regime ($k_B T\ll \hbar^2/ma^2$) the most
important contribution to $g_2$
for hard spheres is
the $s$-wave contribution, which can
be calculated analytically \cite{Larsen}; for non vanishing relative
angular momenta ($l>0$) we neglect the effect of the potential
outside of the hard core.
In this way we can obtain an explicit formula for $g_2$,
\beq
g_{2}(\vec{r},\vec{r}\,';\tau)=1+
\left({\hbar^2\beta\over m}\right)
{1\over rr'}\left\{\exp\left[-{(r+r')^2\over 4\hbar^2\beta/m}\right]
-\exp\left[-{(r+r'-2a)^2\over 4\hbar^2\beta/m}\right]\right\}
e^{-(r r' + \vec{r} \cdot \vec{r}\,')m /2 \hbar^2 \beta}
\marke{swave}
\eeq
for $\vec{r}$ and $\vec{r}\,'$ outside of the hard core diameter
($|\vec{r}|>a$ and $|\vec{r}\,'|>a$), otherwise $g_2=0$.

The density-density correlation function 
can be easily calculated as
\beq
\varrho^{(2)}(\vec{r}\,';\vec{r}\,'',\beta) = 
\sum_{i}\sum_{j\ne i}\langle \delta(\vec{r}\,'-\vec{r}_i)
 \delta(\vec{r}\,''-\vec{r}_j) \rangle.
\marke{dens1} 
\eeq
As the atoms are in a trap rather than in free space,
this quantity is not a function of the relative coordinates
$\vec{r}\,'-\vec{r}\,''$ of the two particles only.
Imagine however that this pair distribution function be probed 
experimentally by scattering of light by the atomic gas, where
we assume a large beam waist compared to the atomic sample. 
As the Doppler effect due to the atomic
motion is negligible, the scattering cross section depends
only on the spatial distribution of the atoms.
Furthermore, for a weak light field very far detuned from the atomic
transitions, the scattering cross section can be calculated
in the Born approximation;  it then depends only on
the distribution function of the relative coordinates 
$\vec{r}\,'-\vec{r}\,''$ between
pairs of atoms.
We therefore take the trace over the center-of-mass position 
$\vec{R}=(\vec{r}\,'+\vec{r}\,'')/2$:
\beq
\varphi^{(2)}(r,\beta) \equiv 
\frac{1}{N(N-1)} \int \, d^3 \vec{R}\,
\varrho^{(2)}(\vec{R}+\vec{r}/2;\vec{R}-\vec{r}/2,\beta),
\marke{rho2}
\eeq
where we have divided by the number of pairs of atoms to normalize
$\varphi^{(2)}$ to unity.
Note that the result depends
only on the modulus $r$ of $\vec{r}$ as the trapping potential
is isotropic.

\subsection{Results of the Simulation}
In fig.1 we show $\varphi^{(2)}(r,\beta)$ for various temperatures
below $T_c^0$, obtained by the simulation of the interacting bosons
in the harmonic trap,
where the critical 
temperature
$T_c$ is reduced compared to the ideal gas
\cite{leggett81,Werner,T_C}.
All pair correlation functions are zero in the region
of the hard-core radius as they should.
At larger length scales the $r$ dependence of the result is also
simple to understand qualitatively, as we discuss now.

Consider first the case $T>T_c$, where no condensate is
present. As the typical interaction energy $n(r)g$ ($n(r)$
being the total one-particle density at $\vec{r}$) is much smaller
than $k_B T$, we expect to recover results close to the
ideal Bose gas. The size of the thermal cloud 
$(k_B T/m \omega)^{1/2}$ determines the
spatial extent of $\varphi^{(2)}(r)$; the bosonic statistics
leads to a spatial bunching of the particles 
with a length scale given by the thermal de Broglie wavelength 
\beq
\lambda_{T}
=\sqrt{\frac{2 \pi \hbar^2}{m k_B T }}.
\eeq
The Bose enhancement of the pair distribution function
is maximal and equal to a factor of 2
for particles at the same
location ($\vec{r}=0$). This effect is preserved by the integration
over the center of mass variable 
and manifests itself through a bump
on $\varphi^{(2)}(r)$ in fig.1. Due to the influence of interactions
the bump is suppressed at small distances and the factor of 2 is
not completely obtained.

For $T<T_c$ a significant fraction of the particles
accumulate in the condensate. As the size of
the condensate is smaller than that of the thermal cloud, 
the contribution to $\varphi^{(2)}$ of the condensed particles
has a smaller spatial extension, giving rise to wings
with two spatial components in $\varphi^{(2)}$, as seen 
in fig.1. Apart from this geometrical effect the building up
of a condensate also affects the bosonic correlations 
at the scale of $\lambda_T$: The bosonic
bunching at this scale no longer exists for particles
in the condensate. This property, referred to as a second order
coherence property of the condensate \cite{ketterle97,burt97,martin},
is well
understood in the limiting case $T=0$; neglecting 
corrections
due to interactions, all the particles are in the
same quantum state $|\psi_0\rangle$ so that e.g.
the 2-body density matrix 
factorizes in a product of one-particle pure state density matrices.
This reveals the absence of spatial correlations between the condensed 
particles.
This explains why in fig.1
the relative height of the exchange bump
with respect to the total height is reduced when $T$ is lowered, that is
when the number of non-condensed particles is decreased.

\section{Comparison with simple approximate treatments}
At this stage a quantitative comparison of the
Quantum Monte Carlo results with well known
approximations can be made.

\subsection{In presence of a significant thermal cloud: 
Hartree-Fock approximation}
As shown in \cite{T_C} in detail, at temperatures sufficiently away from
the critical temperature, the Hartree-Fock approximation
\cite{leggett81} gives
a very good description of the thermodynamic one-particle properties.

To derive the Hartree-Fock Hamiltonian we start from the
second quantized form of the Hamiltonian with contact potential
\beq
\hat{H} = \int \, d^3\vec{r} \,\left[ \hat{\Psi}^{\dagger}(\vec{r})
( H_0 - \mu)
\hat{\Psi}(\vec{r}) \; + \; \frac{g}{2} \,
\hat{\Psi}^{\dagger}(\vec{r}) 
\hat{\Psi}^{\dagger}(\vec{r}) \hat{\Psi}(\vec{r}) 
\hat{\Psi}(\vec{r}) \right]
\eeq
where $H_0$ is the single particle part of the Hamiltonian.
Due to the presence of the condensate we split the field operator
$\hat{\Psi}$ in a classical part $\psi_0$, corresponding to 
the macroscopically occupied ground state and the part of 
the thermal atoms $\hat{\psi}$ with vanishing
expectation value $\langle \hat{\psi} \rangle =0$:
\beq
\hat{\Psi}(\vec{r}) \simeq \psi_0 (\vec{r}) + \hat{\psi} (\vec{r}).
\marke{champ}
\eeq
After this separation we make a ``quadratization'' of the
Hamiltonian by replacing the interaction term by
a sum over all binary contractions of the field operator,
keeping one or two operators
uncontracted, e.g.
\beq
\hat{\psi}^{\dagger} \hat{\psi}^{\dagger} \hat{\psi} \hat{\psi}
\simeq 4 \langle \hat{\psi}^{\dagger} \hat{\psi} \rangle
\hat{\psi}^{\dagger} \hat{\psi} - 2  
\langle \hat{\psi}^{\dagger} \hat{\psi} \rangle
\langle \hat{\psi}^{\dagger} \hat{\psi} \rangle.
\eeq
This is done in such a way
that the mean value of the right hand side agrees with the
mean value of the left hand side in the spirit of 
Wick's theorem.
In the Hartree-Fock approximation we neglect the anomalous
operators, such as $\hat{\psi}^{\dagger} \hat{\psi}^{\dagger}$,
and their averages, and we end up with a Hamiltonian which
is quadratic in $\psi_0$ and $\hat{\psi}$, but also linear in 
$\hat{\psi}$ and $\hat{\psi}^{\dagger}$.
Now we choose $\psi_0$ such that these linear terms vanish
in order to force $\langle \hat{\psi} \rangle=0$.
This gives the Gross-Pitaevskii equation for the condensate
\cite{gpepap}
\beq
\left\{ -\frac{\hbar^2 \nabla^2}{2m}
 + \frac{1}{2}m \omega^2 r^2 + g[n_0(r) +
2 n_T(\vec{r},\vec{r})] \right\} \psi_0(r) = \mu \psi_0(r)
\marke{GP}
\eeq
where 
$n_0(r)=|\psi_0(r)|^2$
corresponds to the condensate density with $N_0$ particles 
and $n_T(\vec{r},\vec{r})=
\langle \hat{\psi}^{\dagger}(\vec{r}) \hat{\psi}(\vec{r}) \rangle$
is the density of the thermal cloud.

Up to a constant term we are left with the Hamiltonian for the
thermal atoms
\beq
\hat{H} = \int \, d^3\vec{r} \, \hat{\psi}^{\dagger}(\vec{r})
( H_0 + 2 g n(r) - \mu )  \hat{\psi}(\vec{r})
\marke{hamhf}
\eeq
where $n(r)=n_0(r) + n_T(\vec{r},\vec{r})$ denotes the total density
and depends only on the modulus of $\vec{r}$.
To work out the density-density correlation function, we formulate
(\ref{dens1}) in second quantization:
\beq
\varrho^{(2)}(\vec{r};\vec{r}\,',\beta) = \langle
\hat{\Psi}^{\dagger}(\vec{r}) \hat{\Psi}^{\dagger}(\vec{r}\,')
\hat{\Psi}(\vec{r}\,') \hat{\Psi}(\vec{r}) \rangle,
\marke{denscor}
\eeq
we use the splitting (\ref{champ}), together with Wick's theorem
and get
\bea
\varrho^{(2)}_{HF}(\vec{r};\vec{r}\,',\beta) & = &
\psi_0(r)\psi_0(r) \psi_0(r') \psi_0(r')
\nonumber \\
&& +
 \psi_0(r)\psi_0(r) n_T(\vec{r}\,',\vec{r}\,')
+ \psi_0(r')\psi_0(r') n_T(\vec{r},\vec{r})
+ 2 \psi_0(r) \psi_0(r')
n_T(\vec{r},\vec{r}\,') 
\nonumber \\
&& + n_T(\vec{r},\vec{r}) n_T(\vec{r}\,',\vec{r}\,') 
+ n_T(\vec{r},\vec{r}\,') n_T(\vec{r},\vec{r}\,').
\marke{Wick}
\eea
Here we have chosen the condensate wave function to be real
and  
\beq
n_T(\vec{r},\vec{r}\,')= \langle
\hat{\psi}^{\dagger}(\vec{r}) \hat{\psi}(\vec{r}\,') \rangle
\eeq
corresponds to the nondiagonal elements of the thermal one body density
matrix.
Since the Hamiltonian (\ref{hamhf}) of the thermal atoms is quadratic
in $\hat{\psi}$,
this density matrix is given by
\beq
n_T(\vec{r},\vec{r}\,') =
\langle \vec{r} | \frac{1}{\exp{ [\beta (H_0 +2 g n(r) - \mu )]} - 1}
| \vec{r}\,' \rangle.
\eeq
In the semiclassical approximation ($k_B T \gg \hbar \omega$)
we can calculate explicitly
these matrix elements by using the Trotter break-up,
which neglects the commutator of $r$ and $p$:
\bea
n_T(\vec{r},\vec{r}\,') & = &
\sum_{l=1}^{\infty}
 \langle  \vec{r}|
e^{ -l \beta (\frac{p^2}{2m} + \frac{1}{2}
m \omega^2 r^2 +2 g n(r) -\mu)} |\vec{r}\,' \rangle \\
& \simeq & \sum_{l=1}^{\infty}
e^{ -\frac{l \beta}{2}
 \left(\frac{1}{2}m \omega^2 r^2 +2 g n(r) -\mu
 \right) }\,
\langle \vec{r} |
e^{ -l \beta \frac{p^2}{2m} } 
| \vec{r}\,' \rangle \,
e^{ -\frac{l \beta}{2}
 \left(\frac{1}{2}m \omega^2 r'^2 +2 g n(r')
-\mu \right) }.
\marke{hf}
\eea
We finally get
\beq
 n_T(\vec{r},\vec{r}\,') = \frac{1}{\lambda_T^3}
\sum_{l=1}^{\infty} \frac{1}{l^{3/2}} \left(
\exp \left[ - \pi \frac{| \vec{r} - \vec{r}\,' |^2}{l \lambda_T^2}
- l \beta \left( m \omega^2 (r^2+r'^2)/4 + g (n(r)+n(r')) - \mu \right)
\right] \right).
\marke{semi}
\eeq
For the diagonal elements the summation gives immediatly the Bose
function $g_{3/2}(z)= \sum_{l=1}^{\infty}
z^l/l^{3/2}$. 
For a given number of particles $N$, eq.(\ref{GP}) and 
the diagonal elements $\vec{r}=\vec{r}\,'$
of eq.(\ref{semi}) have to be solved self consistently
to get the condensate density $n_0(r)$ and the thermal
cloud $n_T(\vec{r},\vec{r})$.
With this solution we can work out the nondiagonal matrix
elements of the density operator which give rise to the
exchange contribution of the density-density correlation
(\ref{Wick}), and the correlation function 
can be written
as a sum over the direct and the exchange contribution
\beq
\varrho^{(2)}_{HF}(\vec{r};\vec{r}\,',\beta)=
\varrho^{(2)}_{direct}(\vec{r};\vec{r}\,',\beta)+
\varrho^{(2)}_{exchange}(\vec{r};\vec{r}\,',\beta).
\eeq

Up to now the short range correlations due to the hard core
repulsion have not been taken into account, but we can improve 
the Hartree-Fock scheme further to include the fact
that it is impossible to find
two atoms at the same location: 
We assume that the particle at $\vec{r}$
interacts with the full Hamiltonian
with the particle at $\vec{r}\,'$
but only with the mean-field of all others
(over which we integrated to derive the reduced density matrix).
This gives in first approximation:
\bea
\widetilde{\varrho}^{(2)}_{HF}(\vec{r};\vec{r}\,',\beta) &=&
\varrho^{(2)}_{direct}(\vec{r};\vec{r}\,',\beta)
g_2(\vec{r}-\vec{r}\,';
\vec{r}-\vec{r}\,',\beta) 
+ \varrho^{(2)}_{exchange}(\vec{r};\vec{r}\,',\beta)
g_2(\vec{r}-\vec{r}\,';\vec{r}\,'-\vec{r},\beta)
\\
&\simeq &
\varrho^{(2)}_{HF}(\vec{r};\vec{r}\,',\beta) g_2(\vec{r}-\vec{r}\,';
\vec{r}-\vec{r}\,',\beta)
\marke{improved}
\eea
where the two particle correlation function $g_2$ 
is the solution of the binary scattering problem, eq.(\ref{swave}).
Further we used the fact that $g_2 \simeq 1$ for particle distances of 
the order of $\lambda_T$ and larger.
In principle one should integrate over the second particle
to get a new one-particle density matrix and find a self-consistent
solution of the Hamiltonian.
But since the range of $g_2$ is of the order of
the thermal wavelength, it will only slightly affect the density, 
so we neglect this iteration procedure.
Using the solution of the coupled Hartree-Fock equations  to 
calculate (\ref{improved}), and
integrating over the center-mass-coordinate, we get
$\varphi_{HF}^{(2)}(r,\beta)$. As shown in fig.1, this 
gives a surprisingly good description of the correlation function
at high and intermediate temperatures.

\begin{figure}
\centerline{ \psfig{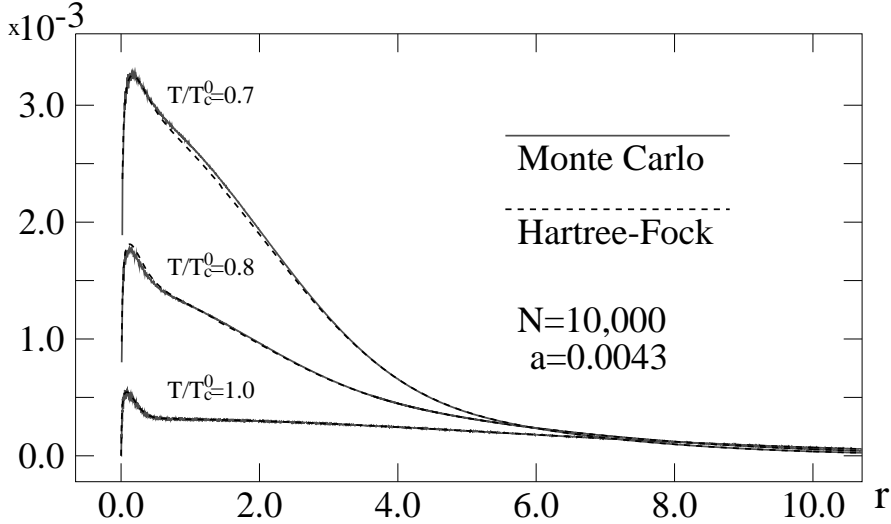} }
\caption{ Pair correlation function $\varphi^{(2)}(r,\beta)$
vs $r$ in units of the harmonic oscillator length
$(\hbar/m \omega)^{1/2}$ 
from  the Monte-Carlo and the Hartree-Fock calculations for
$\beta=0.05 (\hbar \omega)^{-1}$, $\beta=0.06 (\hbar \omega)^{-1}$, and
$\beta=0.07 (\hbar \omega)^{-1}$ (from the bottom to the top). The 
corresponding condensate fractions $N_0/N$ are $0.0$
($T/T_c^0\simeq 1.0$), $0.22$ ($T/T_c^0\simeq 0.8$),
and $0.45$ ($T/T_c^0\simeq 0.7$). $T_c^0$ is the Bose-Einstein
condensation temperature for the ideal gas.
For clarity we removed the part of
short $r$ for the upper curves. }
\end{figure}

\subsection{The quasi-pure condensate: Bogoliubov approach}
The Hartree-Fock description must fail near zero temperature:
Since the anomalous operators $\hat{\psi}^{\dagger} \hat{\psi}^{\dagger}$
and $\hat{\psi} \hat{\psi}$ have been neglected,
it describes not well the low energy excitations of the
systems.
It is known that the zero temperature behavior can be well described
by the Bogoliubov approximation
\cite{bogoliubov47}.
In this paper it is not our purpose to calculate the
correlation function using  the complete
Bogoliubov approach in the inhomogeneous
trap potential. This could be performed using approaches 
developed in \cite{wu96,csordas98}. 
Here we use 
the homogeneous description of the Bogoliubov approximation
and adapt it to the inhomogeneous trap case with a local density
approximation.
This approach already includes the essential features which
the Hartree-Fock description neglects at low
temperatures.
  
We start with the description of the homogeneous system with
quantization volume $V$ and uniform density $n=N/V$. As in \cite{Yvan}
we split the field operator $\hat{\Psi}$ into a 
macroscopically populated state $\Phi$ and a remainder,
which accounts for the noncondensed particles:
\beq
\hat{\Psi}(r)=\Phi(r)\hat{a}_{\Phi} + \delta\hat{\Psi}(r).
\marke{exp}
\eeq
In the thermodynamic limit $N \rightarrow \infty$,
$V \rightarrow \infty$, keeping $N/V=n$ and $N g=const$,
the typical matrix elements of $ \delta\hat{\Psi} $ at low temperatures
are $\sqrt{N}$
times
smaller than $\hat{a}_{\Phi}$. Hence we can neglect
terms cubic and quartic in $ \delta\hat{\Psi}$, when we insert
(\ref{exp}) in the expression of the density-density correlation
function (\ref{denscor}).
Since the condensate density is given by the total
density minus the density of the excited atoms,
we have to express the operator of the condensate density
in the same order of approximation for consistency,
\beq
\hat{a}_{\Phi}^{\dagger}\hat{a}_{\Phi}=\hat{N} - \frac{1}{N}
\int \, d^3 r \, \delta\hat{\Psi}^{\dagger}(r)\hat{a}_{\Phi}
\hat{a}_{\Phi}^{\dagger}\delta\hat{\Psi}(r) + O(N^{-1/2}).
\eeq
Finally we use the mode decomposition of the homogeneous system
\beq
\frac{1}{\sqrt{N}}\hat{a}_{\Phi}^{\dagger}\delta\hat{\Psi}(r)
= \frac{1}{\sqrt{V}}\sum_{\vec{k} \neq 0}
 \left[ \hat{b}_{\vec{k}} e^{i \vec{k} \cdot \vec{r}} u_k + 
\hat{b}^{\dagger}_{\vec{k}} e^{-i \vec{k} \cdot \vec{r}} v_k^* \right]
\eeq
where 
$\hat{b}_{\vec{k}}$ annihilates a quasiparticle with momentum $\vec{k}$.
The components $u_k$ and $v_k$ satisfy the following equations:
\bea
 \left( \begin{array}{cc} \frac{\hbar^2 k^2}{2 m} + g n & g n \\
-g n & - \left( \frac{\hbar^2 k^2}{2 m}+ g n \right) 
\end{array} \right)  
\left( \begin{array}{c} u_k \\ v_k \end{array} \right) 
= E_k \left( \begin{array}{c} u_k \\ v_k \end{array} \right)
\marke{energy}
\eea
together with the normalization:
\beq
|u_k|^2-|v_k|^2=1.
\eeq
At low temperatures the quasiparticles
have negligible interactions and we can
use Wick's theorem to get the following expression for the
correlation function
\bea
\varrho_{BG}(\vec{r}\,';\vec{r}\,'',\beta)  
 =  n^2 + 2 n \int \, d^3 k \,
e^{i \vec{k} \cdot (\vec{r}\,'-\vec{r}\,'')}
\left[ (u_k^2 + v_k^2
+ 2 u_k v_k ) \langle \hat{b}_{\vec{k}}^{\dagger} \hat{b}_{\vec{k}}
 \rangle
+ v_k^2 + u_k v_k \right] + O (\sqrt{n})
\marke{bog}
\eea
where we used $\Phi(r)=V^{-1/2}$.
The 
quasiparticles obey Bose statistics, so that the mean number of 
quasiparticles with momentum $\vec{k}$ and energy $E_k$ is given by
\beq
\langle \hat{b}_{\vec{k}}^{\dagger} \hat{b}_{\vec{k}} \rangle
= \frac{1}{e^{\beta E_k} -1}.
\eeq

We see from eq.(\ref{bog}) that in the homogeneous system the density-density correlation
function depends only on the relative distance $r=|\vec{r}\,'
- \vec{r}\,''|$. The derivation of the properties of the
pair correlation function is given in the appendix. At $T=0 $
the pair correlation function has the following behavior
\cite{huang,note}
\bea
\varphi_{n=const}^{(2)}(r,T=0) 
\simeq \left\{
\begin{array}{cc}
\frac{1}{V}\left[ 1 - \frac{2 a}{r} \right] \simeq
\frac{1}{V}\left[ 1 - \frac{a}{r} \right]^2 & (r \ll \xi) \\
& \\
\frac{1}{V}\left[ 1 - 16 \sqrt{\frac{n a^3}{\pi}}
\left( \frac{\xi}{r} \right)^4
\right] & (r \gg \xi)
\end{array}
\right.
\marke{range}
\eea
where $\xi=(8 \pi n a)^{-1/2}$ is the healing
length of the condensate. For finite but small
temperatures this structure is only slightly changed (see appendix).
The modification of the
low energy spectrum due to the Bogoliubov approach
is responsible for the long range part of the
correlation function.

Apart from the edge of the condensate, the total density
$n(r)$ for low temperature in the trapped system
varies rather slowly compared to the healing
length $\xi$ for the considered parameters. So it is possible to adapt
the result of the homogeneous system to the inhomogeneous
trap case.
For a given density $n(r)$ we get with a local
density approximation 
for the pair correlation function instead of eq.(\ref{bog})
\bea
\varphi_{BG}^{(2)}(r,\beta) & \simeq &
\int \, d^3\vec{R} \,\left\{
n(|\vec{R}+\vec{r}/2|)n(|\vec{R}-\vec{r}/2|) 
\right.
\nonumber \\
&& + 2 n(R) \left.
\int \, d^3 k \, e^{i \vec{k} \cdot \vec{r}} \left[
 (u_k^2(R)  + v_k^2(R)
+ 2 u_k(R) v_k(R))
\left(
e^{\beta E_k(R)} - 1 \right)^{-1}
+ v_k^2(R) + u_k(R) v_k(R) \right] \right\}
\eea
where $u_k(R)$, $v_k(R)$, and 
$E_k(R)$ are solutions of eq.(\ref{energy}) for the given
density $n(R)$.

As shown in fig.2 this gives an excellent agreement with the Quantum 
Monte Carlo results
at low temperature. We have checked that at this temperature the 
difference with the Bogoliubov solution at $T=0$ is almost negligible.
The good agreement with the simulation reflects that the long range
behavior of the pair correlation function in this approximation
is correctly described by eq.(\ref{range}).
We note that in an intermediate
temperature regime, which is not shown,
both approaches, the Hartree-Fock and the local density
Bogoliubov
calculation, do not reproduce the simulation results quantitatively:
The maximum local error is about $5\%$. 
\begin{figure}
\centerline{ \psfig{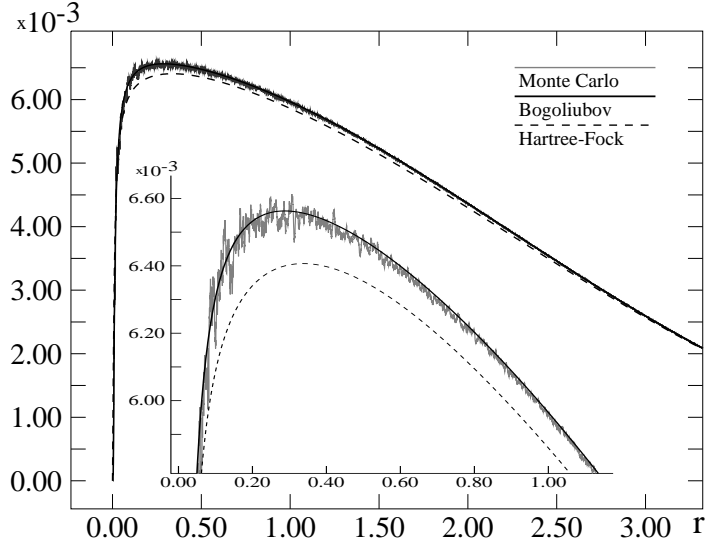} }
\caption{ Pair correlation function 
$\varphi^{(2)}(r,\beta)$ vs $r$ in units of the harmonic oscillator
length $(\hbar/m\omega)^{1/2}$
from the Monte Carlo, the Bogoliubov and the Hartree-Fock
calculations for
$\beta=0.20 (\hbar \omega)^{-1}$ with a condensate fraction
$N_0/N \simeq 0.95$ ($T/T_c^0 \simeq 0.25$).
The healing length is roughly $\xi \simeq 0.3$ in this units }
\end{figure}
\section{Connection to the interaction
energy}
The knowledge of the pair correlation function permits us to
calculate the total energy of the trapped atoms $E_{tot}$:
\beq 
E_{tot}=\frac{1}{\mbox{Tr} \{ \rho \} } \left[
\mbox{Tr}\{H_0 \rho\} + \mbox{Tr}
\{H_I \rho\} \right]
\marke{Etot}
\eeq
One has to pay 
attention that the regularized form of the contact potential,
$V=g \delta(\vec{r})\frac{\partial}{\partial r}
\left( r \cdot \right)$, acts on the off-diagonal elements
$r_{12}$ and $r_{12}'$ of the density-density correlation function
$\varphi(r_{12}',r_{12},\beta)=
\langle \vec{r}_1\,',\vec{r}_2\,' | \rho | \vec{r}_1, \vec{r}_2 \rangle$
in the space of relative coordinates $r_{12}$ and $r_{12}'$.
As the 2-body density matrix
$\varphi(r_{12}',r_{12},\beta)$ diverges as
$(1-a/r_{12}')(1-a/r_{12})$ we actually get the simple form:
\bea
\frac{1}{\mbox{Tr} \{ \rho \}}  \mbox{Tr}
 \{H_I \rho \}  & = &
\frac{N(N-1)g}{4} \int \, d\vec{r}
\frac{\delta(\vec{r})}{r} 
\frac{\partial}{\partial r} \left( r^2
\varphi^{(2)}(r,\beta) \right).
\marke{pseudo}
\eea
This form involves only the diagonal elements of the
correlation function $\varphi^{(2)}(r,\beta)$.
Both the improved Hartree-Fock solution
and the Bogoliubov solution behave for small distances  
($r \ll \xi$) like
\beq
\varphi^{(2)}(r\simeq 0, \beta) \simeq (1 -a/r)^2
\widetilde{\varphi}^{(2)}(0, \beta)
\eeq
where $\widetilde{\varphi}^{(2)}(0, \beta)$
can be obtained graphically by extrapolating the pair correlation
function to zero, neglecting the short range behavior ($r < \xi$);
numerically it can be obtained from the
Hartree-Fock calculation of
(\ref{Wick})
(see \cite{martin} for analysis of the temperature dependence of
$\widetilde{\varphi}^{(2)}(0, \beta)$ ).  
This behavior of the correlation functions shows that eq.(\ref{pseudo})
gives a finite contribution linear in $a$, which we can 
identify with the mean interaction energy $\langle H_I \rangle$:
\beq
\langle H_I \rangle \simeq
g \frac{N(N-1)}{2}  \widetilde{\varphi}^{(2)}(0, \beta).
\marke{Eint}
\eeq
In order $g^2$, eq.(\ref{pseudo}) contains a diverging part,
We note without proof that
this divergency is compensated
within the Bogoliubov theory by a divergent part of the
kinetic energy,
so that the mean total energy, eq.(\ref{Etot}),
is finite. 
This lacks in the Hartree-Fock calculation, which is, however,
limited to linear order of $g$.

In the Thomas-Fermi limit
the kinetic energy is negligible, and the
interaction energy eq.(\ref{Eint}) dominates the 
total energy, which can be measured.
This measurement provides some information about the
correlation function, however, the true correlation
function is not accessible. Only the fictive correlation
function $\widetilde{\varphi}^{(2)}(0, \beta)$ for
vanishing interparticle distances is obtained.

\section{Conclusion}
We numerically calculated the pair correlation function of
a trapped interacting Bose gas with a Quantum Monte Carlo
simulation using parameters typical for recent
experiments of Bose-Einstein
condensation in dilute atomic gases. At temperatures around the
critical point, an improved Hartree-Fock approximation was found 
to be in good quantitative agreement with the Monte Carlo results. 
The improved Hartree-Fock
calculation presented in this paper takes
the short-range behavior of the correlation
function into account, especially the fact that two particles can never
be found at the same location. 
At low temperature we compared our simulation results to a
local density approximation based on the homogeneous Bogoliubov 
approach. The phonon spectrum changes the 
behavior of the pair correlation function for distances $r$
of the order of the healing length $\xi$.
With the knowledge of the pair correlation function
we calculated the total interaction energy.
We showed that the results of recent experiments on second
order coherence do not measure the true correlation
function, which has to vanish for small interparticle distances.
Only an extrapolated correlation function is determined,
where the exact short range behavior disappears.

\section*{Acknowledgments}
This work was partially supported
by the EC (TMR network ERBFMRX-CT96-0002) and
the Deutscher Akade\-mi\-scher Austauschdienst. We are grateful to
Martin Naraschewski, Werner Krauth, Franck Lalo\"e, 
Emmanuel Mandonnet, Ralph Dum and Bart van Tiggelen
for many fruitful discussions.

\section{Appendix}
In this appendix we give the explicit formulas for the pair
correlation function in
the Bogoliubov approach
for an homogeneous system and discuss its behavior 
at short and long distances, since only some aspects have
been discussed in literature \cite{huang,landau}.
 Starting from eq.(\ref{bog}), 
the pair correlation function $\phi^{(2)}_{n=const}$ can be  
be written explicitly as:
\beq
\phi^{(2)}_{n=const}(r,\beta) =
\frac{1}{V} \left[ 1 + \frac{16 a}{\pi r}
\, \int_0^{\infty} \, dq \, \sin(q R) f(q) \right],
\marke{app1}
\eeq
with $R=\sqrt{2}r/\xi$ ($\xi=(8 \pi n a)^{-1/2}$ is the
definition of the healing length) and
\beq
f(q)=\frac{q^2}{\sqrt{1 + q^2}} \left[
e^{\frac{\lambda_T^2}{2 \pi \xi^2} q \sqrt{1 + q^2}} -1 
\right]^{-1} + \frac{q}{2} \left( 
\frac{q}{\sqrt{1+q^2}} - 1 \right).
\marke{app2}
\eeq
To get the behavior of eq.(\ref{app1}) for small distances
($r \ll \xi$), we can replace $f(q)$ by its behavior for large
wavevectors, $q\rightarrow \infty$
\beq
f(q) \sim - \frac{1}{4 q}, \quad q \rightarrow \infty.
\eeq
Using the value of the integral \cite{abramowitz}
\beq
\int_0^{\infty} \, dx \, \frac{\sin x }{x}= \frac{\pi}{2},
\marke{sine}
\eeq
we get the short range behavior of the pair correlation function
\cite{note}:
\beq
\phi^{(2)}_{n=const}(r,\beta)= 
\frac{1}{V} \left[ 1 - 2 \frac{a}{r} \right]
\simeq \frac{1}{V}  \left[ 1 -  \frac{a}{r} \right]^2, \quad r\ll \xi.
\marke{shortrange}
\eeq
To get the long range behavior ($r\gg \xi$), we integrate 
several times by part:
\beq
\int_0^{\infty} \, dq \, \sin(q R) f(q)
= \frac{1}{R} f(0) - \frac{1}{R^3} f^{(2)}(0) + 
\frac{1}{R^5} f^{(4)}(0) - ...  
\eeq
For the function $f(q)$ and its derivatives at $q=0$ we get
\bea
T&=&0: \quad  f(0)=0, \quad f^{(2)}(0)=1 \nonumber \\
T&\neq&0: \quad f(0)=0, \quad f^{(2)}(0)=0, \quad f^{(4)}(0)=0,...
\eea
and the
long range behavior at zero temperature given in (\ref{range})
is obtained.
For finite temperature it can be shown that
$f(q)$ is an odd function of $q$, so that $f^{(2n)}(0)=0$
for all $n$. Due to that the
correlation function vanishes faster than any power law in
$1/R$.

To work out an explicit expression for finite temperatures
we use this antisymmetry to extend
the range of the integral (\ref{app1}) to $-\infty$
and we can analytically calculate the expression for two limiting cases
via the residue calculus. For large distances we only have to take
the poles $q_0$ of $f(q)$ with the smallest modulus into account.
For $\lambda_T/2\pi \ll \xi$ corresponding to $k_B T \gg n g$,
and $r \gg \xi$, we get
$q_0=i$, so that
\beq
\phi^{(2)}_{n=const}(r,\beta)=
\frac{1}{V} \left[ 1 + 2 \frac{1}{n \lambda_T^3}
\frac{\lambda_T}{r}
\exp \left( - \sqrt{2} \frac{r}{\xi} \right) \right].
\eeq
Note the $+$ sign in this expression, leading to $\phi^{(2)}_{n=const}
>1/V$, that we interpret as
a bosonic bunching effect for thermal atoms.
In the opposite limit, $\lambda_T/2 \pi \gg \xi$ and 
$r \gg \lambda_T^2/4 \pi^2 \xi$,
the pole with the smallest
imaginary part is given by $q_0=i 4 \pi^2 \xi^2 / \lambda_T^2$
and we get \cite{landau}
\beq
\phi^{(2)}_{n=const}(r,\beta)=
\frac{1}{V} \left[ 1 - \frac{(2 \pi)^3 }{n}
\frac{4 \pi \xi^4}{\lambda_T^6 r} \exp \left(
- 4 \pi^2 \frac{\sqrt{2} \xi r} {\lambda_T^2} \right) \right].
\eeq


\begin{thebibliography}{99}
\bibitem{anderson95}
M.~H. Anderson, J.~R. Ensher, M.~R. Matthews, C.~E. Wieman,
 and E.~A. Cornell, Science {\bf 269},  198  (1995).


\bibitem{davis95}
K.~B. Davis, M.-O. Mewes, M.~R. Andrews, N.~J. van Druten, D.~S.
Durfee, D.~M. Kurn, and W. Ketterle, Phys. Rev. Lett. {\bf 75},
  3969  (1995).

\bibitem{bradley95} 
C.~C. Bradley, C.~A. Sackett, J.~J. Tolett, and R.~G. Hulet,
Phys. Rev. Lett. {\bf 75}, 1687 (1995);
C.~C. Bradley, C.~A. Sackett, and R.~G. Hulet,
Phys. Rev. Lett. {\bf 78}, 985 (1997).
\bibitem{andrews97} M. R. Andrews, C.G. Townsend, H.-J. Miesner,D.S. Durfee,
D.M. Kurn, and W. Ketterle, Science {\bf 275}, 637 (1997).
\bibitem{roehrl97} A. R\"ohrl, M. Naraschewski, A. Schenzle, and
H. Wallis, Phys. Rev. Lett. {\bf 78}, 4143 (1997).
\bibitem{yang62} C.N. Yang, Rev. Mod. Phys. {\bf 34}, 694 (1962).
\bibitem{ketterle97} W. Ketterle and H.-J. Miesner, Phys. Rev. A {\bf 56},
3291 (1997).
\bibitem{burt97} E.A. Burt, R.W. Ghrist, C.J. Myatt, M.J. Holland,
E.A. Cornell, and C.E. Wieman, Phys. Rev. Lett. {\bf 79}, 337 (1997).

\bibitem{shlyap85} Yu. Kagan, B.V. Svistunov, and G.V.
Shlyapnikov, Pisma. Zh. Eksp. Teor. Fiz. {\bf 42}, 169
(1985) [JETP Lett. {\bf 42}, 209 (1985)].

\bibitem{vanhove54} L. Van Hove, Phys. Rev. {\bf 95}, 249 (1954).

\bibitem{london42} F. London, J. Chem. Phys. {\bf 11}, 203 (1942).

\bibitem{lemmens} F. Brosens, J.T. Devreese, and L. F. Lemmens,
Phys. Rev. E {\bf 55}, 6795 (1997).

\bibitem{martin} M. Naraschewski and R. J. Glauber, preprint 
cond-mat/9806362.


\bibitem{huang} K. Huang, {\em Statistical Mechanics},
(John Wiley \& Sons, MA
1987), chapter 13; T.D. Lee, K. Huang, and C.N. Yang,
Phys. Rev. {\bf 106}, 1135 (1957).

\bibitem{Ceperley} E.L. Pollock and D.M. Ceperley, Phys. Rev. B{\bf 30},
2555 (1984); B{\bf 36}, 8343 (1987);
D.M. Ceperley, Rev. Mod. Phys. {\bf 67},
1601 (1995).

\bibitem{Werner} W. Krauth, Phys. Rev. Lett. {\bf 77}, 3695 (1996).

\bibitem{Feynman} R.P. Feynman, {\em Statistical Mechanics} (Benjamin/
Cummings, Reading, MA, 1972).

\bibitem{barker79} J.A. Barker, J. Chem. Phys. {\bf 70}, 2914 (1979).

\bibitem{Larsen} S.Y. Larsen, J. Chem. Phys. {\bf 48}, 1701 (1968).

\bibitem{leggett81} V.V. Goldman, I.F. Silvera, and A.J. Leggett,
Phys. Rev. B {\bf 24}, 2870 (1981); 
S. Giorgini, L.P. Pitaevskii, and S. Stringari, Phys. Rev. A {\bf 54},
R4633 (1996).

\bibitem{T_C} M. Holzmann, M. Naraschewski,
and W. Krauth, preprint cond-mat/9806201.

\bibitem{gpepap} E.P. Gross, Nuovo Cimento {\bf 20}, 454 (1961);
L.P. Pitaevskii, Sov. Phys. JETP {\bf 13}, 451 (1961).

\bibitem{bogoliubov47} N.N. Bogoliubov, J. Phys. (Moscow) {\bf 11}, 23
(1947).

\bibitem{wu96} A.-C. Wu and A. Griffin, Phys. Rev. A {\bf 54},
4204 (1996).

\bibitem{csordas98} A. Csord\'as, R. Graham, and P. Sz\'epfalusy,
Phys. Rev. A {\bf 57}, 4669 (1998).

\bibitem{Yvan}  C. Gardiner, Phys. Rev. A {\bf 56} 1414 (1997); 
Y. Castin, and R. Dum, Phys. Rev A{\bf 57}, 3008 (1998).

\bibitem{note} For the short range contribution $r \ll \xi$ we get
in our approach $(1-2a/r)/V$  instead of $(1-a/r)^2/V$, which is
the true short range behavior. But the correcting term in
order $a^2$ is of higher order than the calculation. See also 
\cite{huang}.

\bibitem{landau} E.M. Lifshitz and L.P. Pitaevskii, {\em Statatistical
Physics, Part 2} (Pergamon Press, Oxford, 1980), Chapter 9.

\bibitem{abramowitz} M. Abramowitz and I.A. Stegun, {\em
Handbook of mathematical functions}, (Dover Publications, New York,
1972), Chapter 5.

\end{thebibliography}
\end{document}